\documentclass[conference,9pt]{IEEEtran}
\IEEEoverridecommandlockouts
\usepackage{cite}
\usepackage{amsmath,amssymb,amsfonts}
\usepackage{graphicx}
\usepackage{textcomp}
\usepackage{amsmath}
\usepackage{booktabs} 
\usepackage{multirow}
\usepackage{url}            
\usepackage{booktabs}       
\usepackage{amsfonts}       
\usepackage{nicefrac}       
\usepackage{microtype}      
\usepackage{xcolor}         
\usepackage{algorithm}
\usepackage{algpseudocode}
\usepackage{amsmath}
\usepackage{tikz}
\usepackage{stfloats}
\usepackage{orcidlink}
\usepackage{comment}
\usepackage{amssymb}
\usepackage{braket}
\usepackage{xcolor}
\usepackage{subcaption}

\usepackage{enumitem}

\setlist[itemize]{leftmargin=*}%
\setlist[enumerate]{leftmargin=*}%

\usepackage[compact]{titlesec}

\titlespacing\section{0pt}{0.3\baselineskip}{0.2\baselineskip}
\titlespacing\subsection{0pt}{0.2\baselineskip}{0.1\baselineskip}
\titlespacing\subsubsection{0pt}{0.1\baselineskip}{0.1\baselineskip}

\def\BibTeX{{\rm B\kern-.05em{\sc i\kern-.025em b}\kern-.08em
    T\kern-.1667em\lower.7ex\hbox{E}\kern-.125emX}}
\begin{document}

\title{FL-QDSNNs: Federated Learning with Quantum Dynamic Spiking Neural Networks}


\author{\IEEEauthorblockN{Nouhaila Innan\textsuperscript{1,2}, Alberto Marchisio\textsuperscript{1,2}, and Muhammad Shafique\textsuperscript{1,2}
} \IEEEauthorblockA{\textsuperscript{1}eBRAIN Lab, Division of Engineering, New York University Abu Dhabi (NYUAD), Abu Dhabi, UAE\\ \textsuperscript{2}Center for Quantum and Topological Systems (CQTS), NYUAD Research Institute, NYUAD, Abu Dhabi, UAE\\ 
nouhaila.innan@nyu.edu, alberto.marchisio@nyu.edu, muhammad.shafique@nyu.edu\\ }}

\maketitle

\begin{abstract}

We present Federated Learning–Quantum Dynamic Spiking Neural Networks (FL-QDSNNs), a privacy-preserving framework that maintains high predictive accuracy on non-IID client data. Its key innovation is a dynamic-threshold spiking mechanism that triggers quantum gates only when local data drift requires added expressiveness, limiting circuit depth and countering the accuracy loss typical of heterogeneous clients. Evaluated on different benchmark datasets, including Iris, where FL-QDSNNs reach 94\% accuracy, the approach consistently surpasses state-of-the-art quantum-federated baselines; scaling analyses demonstrate that performance remains high as the federation expands to 25 clients, confirming both computational efficiency and collaboration robustness. By uniting adaptive quantum expressiveness with strict data locality, FL-QDSNNs enable regulation-compliant quantum learning for privacy-sensitive sectors and critical infrastructure.

\end{abstract}

\begin{IEEEkeywords}
Quantum Federated Learning, Quantum Spiking Neural Networks, Quantum Parameterized Circuits
\end{IEEEkeywords}

\section{Introduction}
The integration of Quantum Computing (QC) with neural network technologies, including Quantum Machine Learning (QML) \cite{raparthi2022quantum,schuld2014quest,schuld2015introduction,biamonte2017quantum, zaman2023survey}, is a burgeoning field driven by the potential to address specific modern challenges \cite{houssein2022machine,huang2021power,peral2024systematic,abbas2021power,innan2023enhancing,innan2024financial,innan2024financial1,pathak2024resource,innan2024quantum1,kashif2024resqnets,innan2024quantum,dutta2024qadqn,dutta2024aq,innan2025optimizing,innan2025qnn,dave2025sentiqnf, innan2025next, kashif2024computational, siddiqui2024quantum, el2024quantum, marchisio2024cutting, innan2024lep}. Quantum Spiking Neural Networks (QSNNs) combine the dynamic attributes of spiking neural networks, rooted in neurobiological mechanisms, with principles of quantum mechanics \cite{bayro2021quaternion,buscarino2023quaternion}, such as superposition and entanglement \cite{chakraborty2020analytical}, facilitated through quantum gates. At the same time, Federated Learning (FL) is also gaining prominence as a crucial methodology for collaborative model training across distributed nodes \cite{li2020federated,kairouz2021advances, maouaki2025qfal}, effectively addressing significant data privacy and access constraints.

\begin{figure}
    \centering
    \includegraphics[width=1\linewidth]{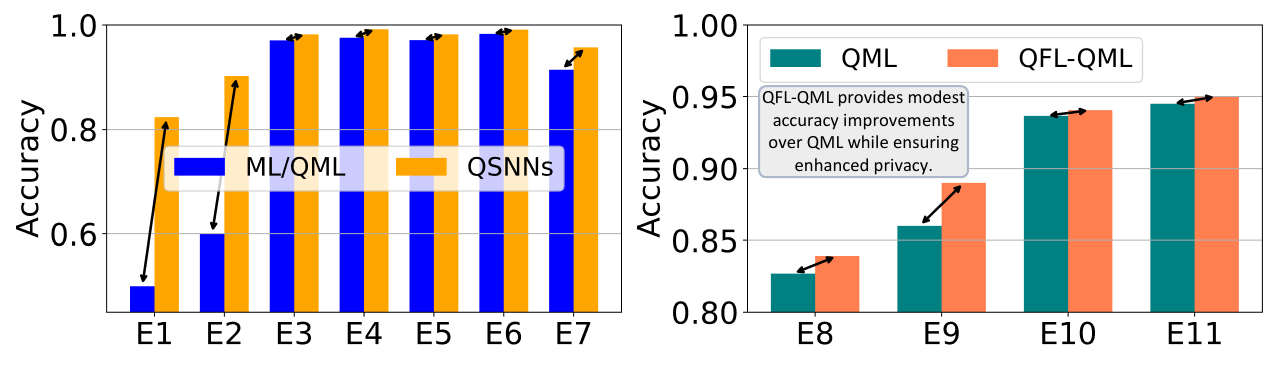}
    \footnotesize
    \setlength{\tabcolsep}{3pt} 
    \renewcommand{\arraystretch}{0.9} 

    \begin{minipage}{0.48\linewidth}
        \centering
\begin{tabular}{|c|c|}
    \hline
    \scriptsize \textbf{Experiment} & \scriptsize \textbf{Dataset [Model]} \\
    \hline
    \scriptsize E1, E2 \cite{sun2021quantum} & \scriptsize MNIST, FMNIST; \\
    & [ANN, QS-SNN] \\
    \scriptsize E3, E4 \cite{xu2024parallel} & \scriptsize MNIST, KMNIST; \\
    & [SFNN, PPF-SQNN] \\
    \scriptsize E5, E6,  & \scriptsize FMNIST, KMNIST, \\
 \scriptsize   E7 \cite{konar2023deep} & \scriptsize CIFAR-10;  \\
 \scriptsize   & [DSNN, DSQ-NET]\\
    \hline
\end{tabular}

    \end{minipage}
    \hfill
    \begin{minipage}{0.48\linewidth}
        \centering
\begin{tabular}{|c|c|}
    \hline
    \footnotesize
    \scriptsize
    \textbf{Experiment} & \scriptsize \textbf{Dataset [Model]}  \\
    \hline
\scriptsize E8 \cite{qu2023dtqfl} & \scriptsize Fetal health; [VQNNs, \\
\scriptsize &  VQNNs-QFL] 
  \\
\scriptsize E9 \cite{qu2024qfsm} & \scriptsize EMO-DB;  [QGRU, \\ 
\scriptsize  &  \scriptsize QFL-QMGU] \\
\scriptsize E10 \cite{chen2021federated} & \scriptsize CIFAR; [hybrid  \\     
\scriptsize & \scriptsize non-FL, QML-FL]  \\
\scriptsize E11 \cite{innan2024qfnn} & \scriptsize Bank transactions;  \\    
\scriptsize & \scriptsize [QNN, QFNN]\\
    \hline
\end{tabular}

    \end{minipage}

    \caption{ \footnotesize
Accuracy comparison across different techniques and datasets based on various existing experiments under specific experimental conditions, as shown in tables. The left chart illustrates that QSNNs consistently outperform \textit{certain} ML and QML models, achieving significantly higher accuracy across diverse datasets. It is important to note that this comparison does not imply QSNNs are universally superior to classical models, which lies outside the scope of this paper. The right chart shows that Federated Quantum Machine Learning (QFL-QML) consistently outperforms standalone QML, delivering higher accuracy. While the primary focus of QFL is to address privacy concerns, this demonstrates its effectiveness in enhancing model performance alongside privacy preservation in distributed learning environments. These results motivate further exploration of integrating QSNNs with QFL, particularly given the promising outcomes observed in classical counterparts. }
    \label{motivation}
\end{figure}

Building upon these considerations, Fig.~\ref{motivation} 
illustrates that in specific experimental conditions—such as non-Independent and Identically Distributed (IID) data distributions and constrained model architectures—QSNNs can outperform conventional ML and QML models on datasets like MNIST. While standard ML models often achieve high accuracy under typical conditions, our experiments highlight scenarios where QSNNs demonstrate superior performance due to their ability to handle complex data dynamics and efficiently learn in federated settings with data privacy considerations. These findings suggest potential strengths of integrating QSNNs with FL methodologies to enhance model performance in specific contexts, motivating further exploration in quantum-assisted learning.

In addition to the observed performance gains under specific experimental conditions, our motivation for developing the FL-QDSNNs framework is rooted in addressing critical challenges unique to quantum-assisted learning. Rather than attempting to surpass classical models outright, our approach is designed to offer a viable solution for quantum environments by ensuring robust privacy and effective adaptation to non-IID data. In QC, where hardware is still emerging and managing quantum states poses inherent complexities, preserving data privacy and accommodating dynamic data distributions are paramount. FL, with its decentralized training paradigm, naturally complements QSNNs by facilitating secure data handling and providing a framework for dynamic adaptation. Thus, the FL-QDSNNs framework not only extends the capabilities of QML but also fills a critical gap by addressing privacy concerns and enhancing the network’s resilience to heterogeneous data—a challenge that is particularly critical in quantum settings.

The integration of QSNNs with FL introduces the Federated Learning-Quantum Dynamic Spiking Neural Networks (FL-QDSNNs) framework, a novel quantum-enhanced method to combine privacy-preserving, scalable, and efficient distributed learning, by leveraging the strengths of QSNNs for complex problem-solving and FL's features of decentralization and privacy preservation, promising to address intricate challenges across diverse domains. \textit{To the best of our knowledge, no previous work has integrated these two technologies with a new approach for QSNNs.}
Despite the promising capabilities, the practical deployment of QSNNs faces several challenges:
\begin{itemize}
\item \textbf{Performance Variability:} QSNNs often struggle with maintaining consistent performance across varied or dynamically changing data distributions, a significant issue where the statistical characteristics of data are unpredictable \cite{ahmadi2021qais}.
\item \textbf{Hardware Limitations:} The reliance on nascent quantum hardware limits QSNNs largely to experimental or simulated environments, impeding their widespread application \cite{kristensen2021artificial}.
\item \textbf{Complex Training Protocols:} QSNNs require intricate training protocols to effectively manage quantum states, presenting challenges in scalability and operational efficiency \cite{brand2024quantum}.
\end{itemize}
The FL-QDSNNs framework aims to harness the theoretical benefits of QSNNs within a federated architecture, enhancing scalability and applicability across diverse datasets. By integrating FL's features with QSNNs, we aim to develop more robust, distributed learning systems.

\begin{figure}[!h]
    \centering
    \includegraphics[width=1\linewidth]{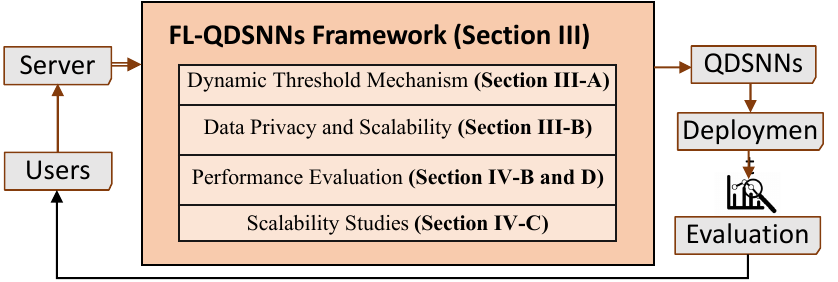}
    \caption{\footnotesize 
    Overview of our novel contributions.}
    \label{novel}
\end{figure}

This exploration establishes a foundational architectural framework for FL-QDSNNs, marking a pivotal advancement in QML. \textbf{The primary contributions of this paper are as follows: (see Fig.~\ref{novel})}
\begin{itemize}
\item \textbf{Innovative Framework:} We develop FL-QDSNNs, a novel framework by seamlessly integrating QSNNs with FL. This strategic combination targets enhanced scalability and robust data privacy in distributed learning environments.
\item \textbf{Dynamic Threshold Mechanism:} Central to our innovation is a novel dynamic threshold mechanism within QSNNs, which adjusts quantum gate activations in response to training dynamics. This mechanism significantly boosts the network's adaptability to diverse and evolving data distributions, directly addressing a key limitation in conventional QSNNs.
\item \textbf{Extensive Validation:} The FL-QDSNNs framework has been rigorously validated across multiple benchmark datasets—specifically Iris, MNIST digits, and breast cancer. Our evaluations include comprehensive ablation studies that highlight the impact of the quantum dynamic spiking mechanism, demonstrating a measurable performance enhancement, including a 3\% increase in accuracy over existing QFL techniques.
\item \textbf{Scalability and Practicality Assessment:} Extensive scalability tests assess the framework's performance with varying numbers of clients, providing actionable insights into optimizing configurations for maximum efficiency and accuracy in practical scenarios.
\end{itemize}

\section{Background and Related Work}
Spiking Neural Networks (SNNs), often referred to as the third generation of neural networks \cite{ghosh2009spiking,tavanaei2019deep, marchisio2023embedded, putra2024embodied, putra2024snn4agents, marchisio2025neuromorphic}, represent an advanced step toward mimicking the actual firing dynamics of biological neurons. Unlike conventional neural networks that process continuous data, SNNs operate on discrete time points where neurons emit spikes in response to stimuli. The fundamental operation within an SNN is the integrate-and-fire model, which captures the essence of neuronal activity and is governed by the following differential equation:
\begin{equation}
\tau \frac{dV(t)}{dt} = -\left(V(t) - V_{\text{rest}}\right) + R \cdot I(t),
\end{equation}
where \( V(t) \) is the membrane potential at time \( t \), \( V_{\text{rest}} \) is the resting membrane potential, \( \tau \) represents the membrane time constant, \( R \) is the membrane resistance, and \( I(t) \) is the input current at time \( t \). A spike is generated when \( V(t) \) exceeds a threshold \( V_{\text{thresh}} \), after which \( V(t) \) is reset. 
This process illustrates how the membrane potential evolves over time in response to input current, leading to spikes.

Recent advancements in neural network technologies have enabled the integration of quantum mechanics, leading to novel computational models such as QSNNs \cite{kristensen2021artificial}. QSNNs have gained attention for their ability to incorporate quantum phenomena, enhancing their robustness in complex data processing scenarios. For instance, QSNNs have been applied to tasks like image background inversion, where conventional ANNs often experience performance degradation due to changes in data statistics \cite{sun2021quantum}. This demonstrated the potential of QSNNs in handling noisy inputs and developing AI systems with brain-like intelligence. Efforts to simulate biological neurons more accurately have also led to the development of Quantum Biological Neurons (QBNs), which use quantum circuits to emulate the integral components of neurons—dendrites, soma, and synapses. This integration of quantum physics with neurobiology offers a richer understanding of neuronal processes by incorporating quantum gates to mimic synaptic functions \cite{lyu2024quantum}. 

Furthermore, the intersection of QC with reinforcement learning has introduced advanced models such as quantum-enhanced Long Short-Term Memory (LSTM) ~\cite{chen2022quantum}. These models mimic the brain's processing functions, integrating sensory data handling with memory functions, similar to the hypothalamus and hippocampus, and have shown effectiveness in reinforcement learning tasks \cite{andres2024brain}. On the hardware front, Single Flux Quantum (SFQ) technology has advanced neuromorphic computing by enabling ultra-fast, energy-efficient data processing. SFQ-based SNN architectures have incorporated spike-timing-dependent plasticity, enabling them to perform complex decision-making tasks effectively \cite{karamuftuoglu2024unsupervised}. Integrating quantum circuits with spiking models has resulted in hybrid systems capable of robust image classification under noisy conditions, showcasing the potential for quantum mechanics and neuromorphic computing to address real-world challenges \cite{konar2023shallow}. These quantum and neural network technological advances promise to revolutionize computational neuroscience, and AI \cite{chen2022accelerating,bayro2021quaternion}.

\section{FL-QDSNNs Framework}
\begin{figure}[htpb]
    \centering
    \vspace{-8pt}
\includegraphics[width=1\linewidth]{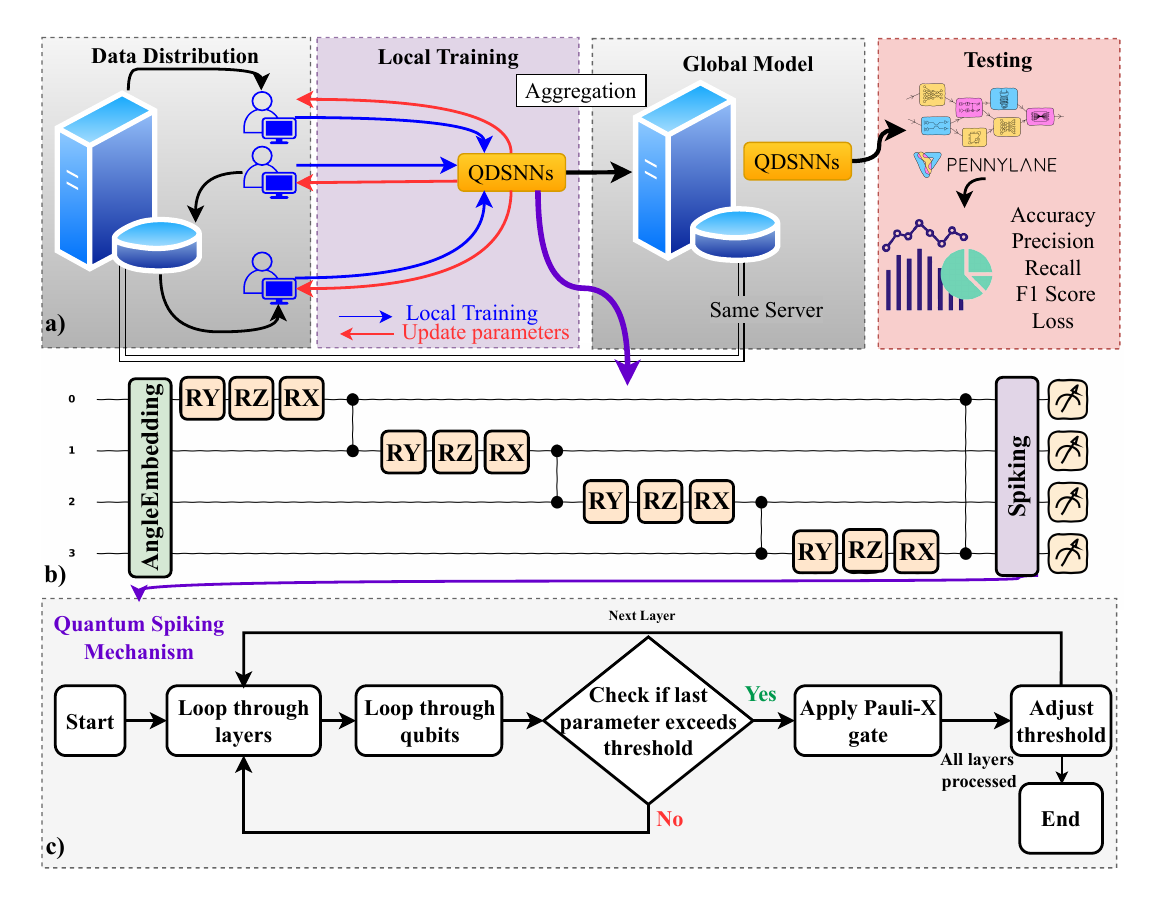}
    \caption{ \footnotesize
 (a) The workflow starts with data distribution from a central server to multiple clients, each employing QDSNNs for local data processing and parameter updating via the Adam optimizer. These parameters are centrally aggregated to refine the global model, which is then processed through additional quantum circuits and evaluated using various metrics via PennyLane. (b) The illustration details a QSNN's single-layer quantum circuit, starting with AngleEmbedding for encoding data into the quantum states of four qubits, followed by parameterized rotations and Controlled-Z gates to promote inter-qubit interactions and enhance quantum correlations. The circuit concludes with a global ``Spiking'' operation, depicting the neural firing mechanism within QDSNNs. (c) The quantum spiking mechanism is showcased, emphasizing the sequential processing of each layer's qubits. It includes checks for whether the last parameter of a qubit exceeds a dynamic threshold, possibly triggering a Pauli-X gate, followed by threshold adjustments before the next layer is processed, completing the network's operation cycle.}
    \label{framework}
\end{figure}

This section outlines the design and operational workflow of the FL-QDSNNs framework. As illustrated in Fig. \ref{framework}, the process begins with data distribution across multiple clients, followed by local quantum-enhanced computations that employ quantum spiking dynamics.
 These computations are then synchronized through the aggregation of parameters at a central server to refine the global model. The detailed training procedure, including data distribution, local model updates, and global optimization, is formalized in Algorithm~\ref{alg1}.
\begin{algorithm}[htpb]
\caption{FL-QDSNNs}
\begin{algorithmic}[1]
\footnotesize
\State \textbf{Input:} Full Dataset $X$ with labels $Y$
\State \textbf{Input:} Corresponding labels $Y_1, Y_2, \ldots, Y_n$
\State \textbf{Input:} Base learning rate $\eta$, dynamic spike threshold $\tau \in [0, 1]$, number of epochs $E$
\State \textbf{Output:} Trained FL-QDSNNs parameters $\theta$
\Procedure{DistributeDataNonIID}{$X, Y, n$}
    \State Partition $X$ and $Y$ into $n$ subsets $\{(X_1, Y_1), \ldots, (X_n, Y_n)\}$
    \Comment{Partition data for $n$ clients}
    \For{each subset $(X_i, Y_i)$}
        \State Adjust the class distribution in $X_i$ to introduce variability
        \Comment{Ensure non-IID characteristics across clients}
    \EndFor
    \State \textbf{return} $\{(X_1, Y_1), \ldots, (X_n, Y_n)\}$
\EndProcedure
\Procedure{TrainQF-QSFNN}{$X, Y, \eta, \tau, E$}
    \State Initialize global parameters $\theta_g \gets \text{random initialization}$
    \For{each epoch $e = 1$ to $E$}
        \For{each client $i = 1$ to $n$}
            \State $\theta_i \gets \theta_g$ \Comment{Distribute global parameters to client $i$}
            \State $\theta_i, L_i$ $\gets$ \Call{LocalUpdate}{$X_i$, $Y_i$, $\theta_i$, $\eta$, $\tau$}
        \EndFor
        \State $\theta_g \gets \frac{1}{n} \sum_{i=1}^n \theta_i$ \Comment{Aggregate updated parameters}
    \EndFor
    \State \textbf{return} $\theta_g$
\EndProcedure

\Procedure{LocalUpdate}{$X, Y, \theta, \eta, \tau$}
    \For{each sample $(x, y) \in (X, Y)$}
        \State $p \gets \Call{QuantumCircuit}{x, \theta, \tau}$ \Comment{Forward pass through QSNN}
        \State $L \gets \Call{Loss}{p, y}$ \Comment{Compute loss}
        \State $\theta \gets \theta - \eta \cdot \nabla_\theta L$ \Comment{Parameter update}
    \EndFor
    \State \textbf{return} $\theta, L$
\EndProcedure

\Procedure{QuantumCircuit}{$x, \theta, \tau$}
    \State \textbf{Input:} Data sample $x$, parameters $\theta$, threshold $\tau$
    \State \textbf{Qubits:} Initialize 4 qubits
    \State Apply angle encoding $RY(\pi \cdot x[q])$ to all the qubits only once
    \For{layer $l = 1$ to $5$}
        \For{qubit $q = 0$ to $3$}
            \State Apply parameterized gate $U(\theta_{l,q,1}, \theta_{l,q,2}, \theta_{l,q,3})$ to qubit $q$
            \State Apply controlled gate $CZ(q, (q+1) \mod 4)$
        \EndFor
    \EndFor
    \For{each qubit $q$}
        \If{$\theta_{l,q,n} > \tau$} \Comment{Check last layer's parameters for each qubit}
            \State Apply $PauliX$ to qubit $q$ \Comment{Activate spike if above threshold}
        \EndIf
    \EndFor
    \State $p \gets \Call{MeasureAllQubits}{}$ \Comment{Measure all qubits to get collective output}

    \State \textbf{return} $p$
\EndProcedure
\end{algorithmic}
\label{alg1}
\end{algorithm}
\subsection{Data Distribution} In our FL-QDSNNs framework, data distribution plays a crucial role in modeling realistic FL scenarios. We distribute the datasets across multiple clients in a non-IID manner. This approach ensures that each client has a unique subset of data, with variations in class distributions that reflect the heterogeneous nature of real-world data environments.
To achieve non-IID distribution, we partition the data in such a way that the subsets from different clients are imbalanced or biased towards certain classes. For example: 
\begin{itemize} 
\item \textbf{Class Imbalance}: Certain clients receive data heavily skewed toward specific classes, while others may have a more balanced representation. 
\item \textbf{Feature Variability}: In some cases, clients' datasets exhibit feature variations representing distinct environments or conditions under which the data was collected. 
\end{itemize}

This non-IID setup is critical for testing the robustness and adaptability of the FL-QDSNNs framework. By simulating data heterogeneity, we can better evaluate the framework's ability to generalize across diverse data distributions while maintaining high performance in an FL context.
\subsection{Quantum Parameterized Circuit and Quantum Dynamic Spiking Mechanism}
QDSNNs within our FL-QDSNNs framework blend biological neural network dynamics with the advanced capabilities of QC, significantly boosting both computational power and data privacy in a quantum-enhanced FL environment. QDSNNs employ a sophisticated quantum-parameterized circuit that starts with encoding classical data into quantum states like any QML models—critical for adapting classical information for quantum manipulation:
\begin{equation}
\ket{\psi_{\text{in}}} = \bigotimes_{j=1}^n RY(x_j)(\ket{0}),
\end{equation}
where \(RY(x_j)\) is a rotation gate that encodes the data feature \(x_j\) onto the \(j\)-th qubit, preparing the quantum state for processing. This initial encoding is essential for representing classical data within the quantum framework, setting the stage for complex quantum operations.

In this circuit, QDSNNs process the data through multiple layers of parameterized gates, which are crucial for the network's ability to learn and adapt:
\begin{equation}
\ket{\psi_{\text{out}}} = U(\theta) \ket{\psi_{\text{in}}},
\end{equation}
where \(U(\theta)\) comprises unitary operations parameterized by the vector \(\theta\), including both single-qubit rotations and controlled entangling operations. These operations enable the QDSNNs to explore the high-dimensional landscape of quantum states (see Fig. \ref{framework}-b).

\textit{Central to the QDSNNs circuit is a parameter-dependent spiking mechanism where the decision to apply the Pauli-X gate (simulating the spiking action) is based on the magnitude of the last parameter in the sequence of parameterized gates within each layer.} This approach is inspired by how neurons in biological neural networks fire an action potential when their membrane potential reaches a threshold. 

In QDSNNs, we draw a parallel by treating the last parameter of each qubit in each layer as a proxy for this potential:
\begin{equation}
\text{if } \theta_{l,q,n} > \tau, \text{ apply } X_q,
\end{equation}
where n is the last parameter. 
This conditional application of the Pauli-X gate based on the last parameter of each layer provides several advantages:
\begin{itemize}

\item \textbf{Layer-Specific Adaptivity:} It allows each layer to adaptively respond based on its own computational conclusions, thus incorporating a form of depth-wise learning where deeper layers can contribute differently to the decision-making process based on more complex transformations of the input data.

\item \textbf{Temporal Dynamics:} By using the last parameter, the QDSNNs can incorporate a form of temporal dynamic into the learning process, akin to the time-varying nature of neuronal action potentials. This can be particularly beneficial in tasks that require memory or statefulness over sequential data inputs.

\item \textbf{Enhanced Non-Linearity:} Applying a non-linear operation like the Pauli-X gate based on a threshold condition helps introduce additional non-linearity into the quantum circuit, which is crucial for complex problem-solving capabilities in QML.
\end{itemize}

The threshold \( \tau \) itself is not static but is dynamically adjusted during the learning process to optimize the responsiveness of the QDSNNs
$\tau \gets \tau + \Delta \tau$ \text{ (adjustment based on network performance metrics)},
where \(\Delta \tau\) represents the incremental changes made to the threshold, which are guided by the overall network performance metrics such as accuracy or loss. This dynamic adjustment helps the network maintain an optimal balance between sensitivity (ability to detect changes) and specificity (ability to ignore non-essential variations).

The final measurement of the quantum state collapses the state into classical outputs \(p\), which are utilized to compute the QDSNNs output and the subsequent loss for backpropagation. This unique spiking mechanism mimics biological processes and employs quantum mechanical principles to enhance computational capabilities while preserving data privacy across the federated network.

\subsection{Privacy, Training, and Optimization}
The process of training our FL-QDSNNs framework begins with the initialization of global parameters \(\theta_g\), which are crucial for setting the initial conditions of the quantum parameterized circuits utilized by each client. These parameters are distributed to all clients, ensuring that each node starts the learning process with a uniform set of parameters, thereby promoting consistent learning across the network.

Each client conducts local updates by employing the QDSNNs circuit designed to effectively process quantum data. The local update rule is rigorously defined as:
\begin{equation}
\theta_i^{(t+1)} = \theta_i^{(t)} - \eta \nabla_{\theta_i} L(X_i, Y_i, \theta_i^{(t)}),
\end{equation}
where \(\eta\) denotes the learning rate, and \(\nabla_{\theta_i} L\) represents the gradient of the loss function \(L\), which is calculated using the backpropagation through time method adapted for quantum parameterized circuits.

While various specific loss functions exist, for this particular architecture, we employ a classical loss function that has been successfully applied in similar settings. Specifically, we use the Mean Squared Error (MSE) for classification tasks, defined as:
\begin{equation}
L = \frac{1}{N} \sum_{n=1}^N (y_n - \hat{y}_n)^2,
\end{equation}
where \(y_n\) is the true label, and \(\hat{y}_n\) is the predicted label obtained from the quantum measurement process. When it comes to optimizing this loss function and updating the parameters after each iteration, the Adam optimizer is utilized due to its efficiency in handling the non-convex optimization landscapes typical in quantum parameter spaces:
\begin{equation}
\theta_{t+1} = \theta_t + \text{Adam}(\nabla_{\theta} L),
\end{equation}

Post-local optimization, parameters are aggregated to update the global model:
\begin{equation}
\theta_g^{(t+1)} = \frac{1}{n} \sum_{i=1}^n \theta_i^{(t+1)}.
\end{equation}
This averaging method ensures that each client's updates contribute equally to the global model, enhancing the overall system's ability to generalize across diverse datasets.

\subsection{Evaluation and Performance Metrics}
The FL-QDSNNs framework is rigorously evaluated using a suite of metrics to assess its performance and scalability across diverse scenarios. We measure accuracy, MSE, and generate detailed classification reports to evaluate model effectiveness and error characteristics. Our analysis includes studying training dynamics by tracking accuracy and loss over iterations to understand learning behavior, examining the impact of varying client numbers on global accuracy to assess scalability, and exploring threshold sensitivity to optimize the FL-QDSNNs response to data variability. This comprehensive evaluation helps us refine the framework's capabilities, ensuring robustness and adaptability in the FL environment.

\section{Experiments}
\subsection{Experimental Setup}
Following the configuration and hyperparameters outlined in Table \ref{hyper}, our FL-QDSNNs framework is rigorously tested across three datasets: Iris, digits, and breast cancer. The data from each dataset is distributed among clients in a non-IID manner, meaning that each client receives a subset of the data with variations in class distributions, simulating realistic FL scenarios. The setup involved a diverse number of clients and a thorough exploration of quantum capabilities through the Pennylane default simulator \cite{pennylane}. 
\begin{table}[htpb]
\centering
\renewcommand{\arraystretch}{1.2} 
\caption{Configuration and hyperparameters of FL-QDSNNs}
\label{hyper}
\begin{tabular}{@{}p{0.4\linewidth}p{0.5\linewidth}@{}} 
\toprule
\textbf{Parameter} & \textbf{Value} \\
\midrule
Datasets  & Iris \cite{iris_53}, digits\cite{digits_80}, breast cancer\cite{breast_cancer_14} \\ 
Parameters & 12 per layer, randomly initialized\\ 
Number of qubits and layers & 4 and 5 \\ 
Quantum device & Pennylane default simulator (default.qubit)   \\ 
Optimizer  & Adam \\ 
Stepsize for optimizer & 0.05\\ 
Local iterations per client & 100\\ 
Global iterations & 100\\ 
Dynamic spike threshold & Increment by 0.05 per iteration \\ 
Number of data points & Iris: 150, digits: 1,797, breast cancer: 569 \\ 
Features used & 4 (reduced features for digits and breast cancer)  \\                    
Dataset split (each dataset) & 80\% training, 20\% testing \\ 
Number of clients & 5, 10, 15, \textbf{20}, 25 \\ 
Evaluation metrics & Accuracy, MSE, classification report\\ 
\bottomrule 
\end{tabular}
\end{table}

\subsection{Training Dynamics: Accuracy and Loss}
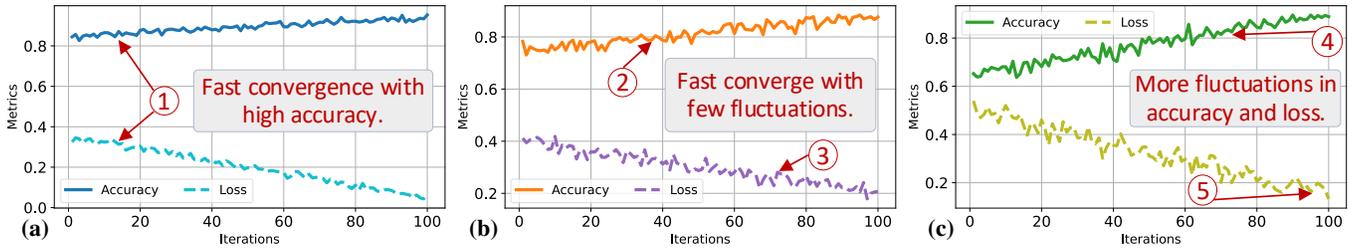
\begin{figure*}[h!]
    \centering
    \begin{tikzpicture}
        \node[anchor=south west,inner sep=0] (image) at (0,0) {\includegraphics[width=1\linewidth]{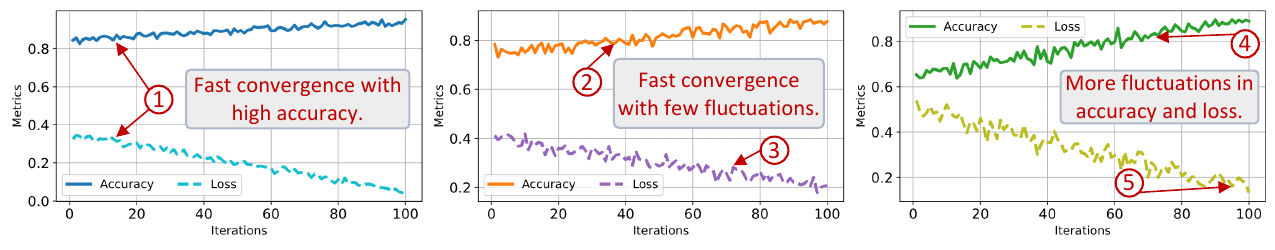}};
        \node[anchor=south west, font=\bfseries\small, text=black] at (0.25,0.1) {(a)};
        \node[anchor=south west, font=\bfseries\small, text=black] at (6.2,0.1) {(b)};
        \node[anchor=south west, font=\bfseries\small, text=black] at (12.3,0.1) {(c)};    
    \end{tikzpicture}
        \caption{\footnotesize
 Training performance for the three benchmark datasets over 100 iterations: \textbf{(a)} Iris, \textbf{(b)} Digits, and \textbf{(c)} Breast cancer. Key moments in the training process are marked: \textcircled{\raisebox{-0.9pt}{1}} denotes the rapid learning phase in the Iris dataset; \textcircled{\raisebox{-0.9pt}{2}} and \textcircled{\raisebox{-0.9pt}{3}} emphasize a notable increase in accuracy and a consistent decline in loss, respectively, for the digits dataset. \textcircled{\raisebox{-0.9pt}{4}} highlights significant fluctuations in the accuracy, and \textcircled{\raisebox{-0.9pt}{5}} marks the point of minimal loss in the breast cancer dataset.}
     \label{res}

\end{figure*}
As shown in Fig. \ref{res}, the training dynamics of our FL-QDSNNs across different datasets reveal distinct learning behaviors and adaptations. For the Iris dataset, the significant early improvement marked by \textcircled{\raisebox{-0.9pt}{1}} within the initial 10 iterations highlights the FL-QDSNNs' swift and effective response to the dataset's simpler structure. This rapid phase of learning, supported by the spiking mechanisms, showcases the network's ability to rapidly converge to high accuracy and low loss, reflecting efficient FL dynamics.
For the digits dataset, the markers \textcircled{\raisebox{-0.9pt}{2}} and \textcircled{\raisebox{-0.9pt}{3}} illustrate a more nuanced learning curve in the federated context. At iteration 20, the marked increase in accuracy (\textcircled{\raisebox{-0.9pt}{2}}) demonstrates our framework's capability to adapt to complex numerical data, showing that spiking dynamics potentially enhance computational efficiency and learning depth. The consistent decrease in the loss by iteration 50 (\textcircled{\raisebox{-0.9pt}{3}}) further verifies the effective federated adaptation of our framework, showing the network's ability to generalize well across more complex data without memorizing.
In the training curve for the breast cancer dataset, \textcircled{\raisebox{-0.9pt}{4}} and \textcircled{\raisebox{-0.9pt}{5}} mark critical stages of learning in a highly variable medical context. The fluctuations in accuracy at \textcircled{\raisebox{-0.9pt}{4}} reveal challenges in FL from diverse and complex data sources, reflecting our framework's ongoing adjustments to optimize learning. Nonetheless, achieving minimal loss at \textcircled{\raisebox{-0.9pt}{5}} by iteration 80 illustrates our framework's robust learning capabilities, effectively minimizing errors and demonstrating the potential of our FL-QDSNNs framework for handling intricate datasets.
\begin{figure*}[b]
\centering
    \begin{tikzpicture}
        \node[anchor=south west,inner sep=0] (image) at (0,0) {\includegraphics[width=1\linewidth]{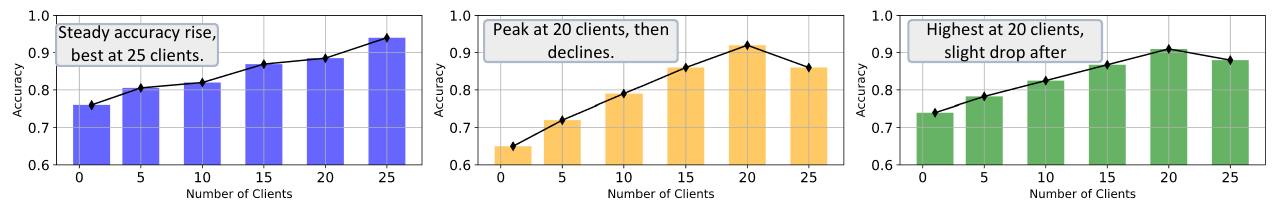}};
        \node[anchor=south west, font=\bfseries\small, text=black] at (0.25,0.1) {(a)};
        \node[anchor=south west, font=\bfseries\small, text=black] at (6.2,0.1) {(b)};
        \node[anchor=south west, font=\bfseries\small, text=black] at (12.3,0.1) {(c)}; 
                    \begin{scope}[x={(image.south east)},y={(image.north west)}]
                \node[circle, draw=blue, fill=white, inner sep=1pt] (pointerC) at (0.19,0.85) {\textcolor{blue}{1}};
                \draw[-latex, blue, thick] (pointerC) -- +(0.06, -0.1);
            \end{scope}
            \begin{scope}[x={(image.south east)},y={(image.north west)}]
                \node[circle, draw=blue, fill=white, inner sep=1pt] (pointerC) at (0.55,0.87) {\textcolor{blue}{2}};
                \draw[-latex, blue, thick] (pointerC) -- +(0.038, -0.08);
            \end{scope}  
            \begin{scope}[x={(image.south east)},y={(image.north west)}]
                \node[circle, draw=blue, fill=white, inner sep=1pt] (pointerC) at (0.88,0.87) {\textcolor{blue}{3}};
                \draw[-latex, blue, thick] (pointerC) -- +(0.038, -0.1);
            \end{scope}             
    \end{tikzpicture}
 \caption{\footnotesize
 Impact of client counts on accuracy for the three datasets: \textbf{(a)} Iris, \textbf{(b)} Digits, and \textbf{(c)} Breast Cancer. Each subplot presents accuracy trends as a function of increasing client numbers, depicted by bars connected by lines. Noteworthy points include: \textcircled{\raisebox{-0.9pt}{1}} showing a steady accuracy increase with more clients for Iris, \textcircled{\raisebox{-0.9pt}{2}} identifying peak accuracy at 20 clients for digits followed by a decline, and \textcircled{\raisebox{-0.9pt}{3}} displaying a similar trend for breast cancer, with accuracy peaking before decreasing at higher client counts.}
\label{acc-clients}
\end{figure*}
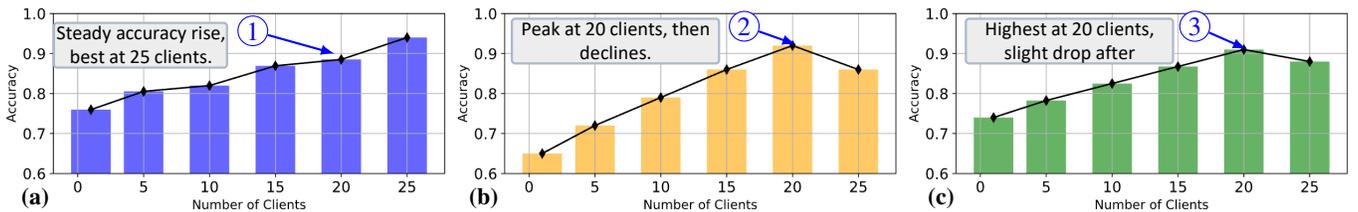

\subsection{Scalability Effects: Client Number Impact on Global Accuracy}

The relationship between the number of clients and the global accuracy across three datasets in the FL context is presented in Fig. \ref{acc-clients}. For the Iris dataset (\textcircled{\raisebox{-0.9pt}{1}}), a continuous increase in accuracy is observed as more clients participate. This trend suggests that our FL-QDSNNs framework effectively utilizes the distributed data, likely due to the simpler nature of the dataset, which facilitates efficient generalization across a wider array of data contributions.
For the digits dataset, the accuracy trend demonstrates significant improvements as the number of clients increases, reaching a peak at 20 clients (\textcircled{\raisebox{-0.9pt}{2}}). Beyond this optimal client count, the accuracy slightly declines, possibly indicating that while the FL framework scales well to a certain extent, an excessive number of clients might introduce conflicting data patterns or noise, which marginally degrades our FL-QDSNNs framework performance.
Similarly, the breast cancer dataset shows an increase in accuracy with the number of clients up to 20 (\textcircled{\raisebox{-0.9pt}{3}}), after which a slight reduction in accuracy is noted. This pattern suggests that although our FL-QDSNNs framework initially benefits from the additional data, contributing to enhanced learning, it eventually faces diminishing returns or complications from overfitting as the complexity and variability of data from an excessive number of sources influence the learning process.
These observations underscore the complex impact of client diversity and quantity in FL setups. They reveal that while increasing the number of clients typically benefits training, an optimal client threshold exists beyond which additional contributions may not enhance and could potentially impair accuracy.

\subsection{Threshold Sensitivity: Adjusting Spike Dynamics for Optimal Accuracy}

\begin{figure}[h!]
    \centering
    \includegraphics[width=1\linewidth]{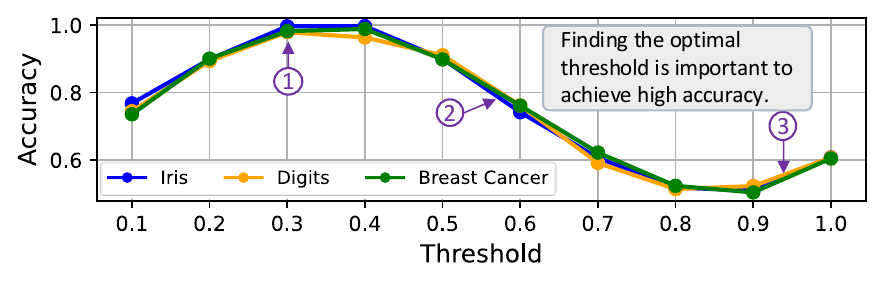}
    \caption{
 Impact of the spike threshold on accuracy across the three datasets: Iris, Digits, and Breast Cancer. Each line tracks the accuracy changes with adjustments in threshold levels. Key observations are noted: \textcircled{\raisebox{-0.9pt}{1}} identifies the peak accuracy point across all datasets, indicating the optimal threshold for maximal accuracy; \textcircled{\raisebox{-0.9pt}{2}} marks a subsequent decline in accuracy, highlighting the threshold sensitivity of the FL-QDSNNs; \textcircled{\raisebox{-0.9pt}{3}} shows a recovery in accuracy, underscoring the non-linear effects of threshold adjustment on FL-QDSNNs.}
    \label{acc-spiking}
\end{figure}

We examine how adjustments in the spiking threshold influence the accuracy of our framework across the three datasets, as described in Fig. \ref{acc-spiking}. The graph provides a clear visual representation of the critical points where changes in threshold settings affect the FL-QDSNNs performance. The peak accuracy, marked by \textcircled{\raisebox{-0.9pt}{1}}, represents the optimal threshold setting across all datasets, suggesting that a moderate threshold maximizes the extraction of useful signals while minimizing noise for the most effective learning. Following this peak, a noticeable decline in accuracy is observed, highlighted by \textcircled{\raisebox{-0.9pt}{2}}, indicating the threshold limit beyond which our framework generalization capabilities begin to deteriorate rapidly. This is attributed to over-suppression of neuron firing, which may discard valuable information essential for accurate predictions. 

However, as the threshold approaches its maximum, a slight recovery in accuracy is seen, as denoted by \textcircled{\raisebox{-0.9pt}{3}}. This suggests that while our FL-QDSNNs framework becomes overly restrictive at high thresholds, it still captures some relevant input data features. However, the overall accuracy remains lower compared to the peak. This behavior highlights the non-linear and complex nature of threshold adjustments in FL-QDSNNs, underscoring the necessity of carefully selecting spiking thresholds to optimize performance by aligning with the dataset's specific characteristics and noise levels.

\subsection{Comprehensive Evaluation and Comparison} 
The performance metrics, summarized in Table \ref{tab:dataset_metrics}, demonstrate the efficacy of our framework. For the Iris dataset, our framework achieves an accuracy of 94\%, significantly outperforming existing QFL frameworks. In particular, our approach surpasses the FedQNN model, which reports an accuracy of 90\% \cite{innan2024fedqnn}, and the QFNN model, which achieves an accuracy of 64\% \cite{gurung2024personalized}. These comparisons underscore the advancements in our approach, leveraging optimized quantum circuits through the quantum spiking mechanism and efficient FL protocols that enhance both computational efficiency and learning accuracy.

Likewise, the results for the digits and breast cancer datasets suggest that our approach is robust across various types of data, manifesting high recall, precision, and F1 scores, critical for practical applications. The adaptability and scalability of our framework propose significant potential for broader applications in QML, especially in scenarios requiring stringent privacy and data security, as ensured by our framework.
\begin{table}[htpb]
\footnotesize
\centering
\caption{Testing performance metrics for the Iris, digits, and breast cancer datasets.}
\label{tab:dataset_metrics}
\begin{tabular}{@{}lcccc@{}}
\toprule
\textbf{Dataset}        & \textbf{Accuracy} & \textbf{Recall} & \textbf{Precision} & \textbf{F1 Score} \\ \midrule
Iris           & 0.94     & 0.93   & 0.92      & 0.92   \\
Digits         & 0.89     & 0.87   & 0.88      & 0.87    \\
Breast Cancer  & 0.91     & 0.90   & 0.92      & 0.91    \\ \bottomrule
\end{tabular}
\end{table}
\subsection{Ablation Study: Impact of Quantum Dynamic Spiking}
To highlight the impact of the quantum dynamic spiking mechanism within our FL-QDSNN framework, we conduct an ablation study by deactivating the dynamic threshold-based spiking function and observing performance changes across various datasets distributed among clients with the same settings. 
This allowed us to isolate the specific contributions of this mechanism from the overall framework, examining performance changes across three datasets distributed among clients. 

Our findings show that excluding this mechanism leads to a significant performance decline. On the Iris dataset, accuracy drops from 94\% to 86\%, indicating the mechanism's effectiveness in managing simpler structures. The digits dataset experiences a reduction from 89\% to 82\%, emphasizing the mechanism's critical role in processing complex numerical data. For the breast cancer dataset, accuracy decreases from 91\% to 84\%, underscoring its significance in navigating complex medical data scenarios.
By dynamically adjusting quantum gate activation thresholds in response to real-time data variations and distribution, the mechanism substantially enhances the framework's robustness. This adaptability allows the model to capture critical data features effectively, improving generalization across diverse and non-IID datasets in an FL environment.
\section{Discussion and Conclusion}
The outcomes of our study underscore the potential of FL-QDSNNs to effectively handle diverse and non-IID datasets. By achieving high performance metrics, our framework demonstrates its robustness and adaptability. These attributes are crucial for applications in fields where accuracy, privacy, and security are paramount. Our comprehensive ablation studies further validate the impact of the quantum dynamic spiking mechanism, revealing how its absence leads to significant performance declines and highlighting the framework's ability to generalize effectively under non-IID conditions.

One of the more significant insights from this study is identifying an optimal client threshold in FL setups. This finding suggests that while adding more nodes to the network generally improves model performance due to increased data diversity, there is a point beyond which additional data may introduce diminishing returns or even degrade performance due to noise or conflicting data patterns. This observation is critical for deploying FL systems, particularly in environments with variable data quality or highly heterogeneous data sources.
Furthermore, the sensitivity of the model performance to the spiking threshold adjustments highlights the importance of parameter tuning in QDSNNs. The optimal balance between neuron firing suppression and signal extraction is key to maximizing model accuracy and generalization. This balance may vary by dataset but also by the specific characteristics of the data, such as feature distribution and noise levels.

Looking forward, the promising results obtained here lay the groundwork for future explorations into more complex quantum circuit designs and the integration of advanced QC techniques into FL frameworks. Further research could explore the impacts of different quantum gate configurations, deeper circuit architectures, or hybrid quantum-classical models to enhance learning efficiency and scalability. In summary, the study opens avenues for substantial advancements in QML, particularly in optimizing FL frameworks to harness the unique capabilities of QC and QML for real-world applications. These advancements could significantly impact sectors like healthcare, finance, and cybersecurity, where data privacy is crucial, and the stakes for accuracy and reliability are high.
\section*{Acknowledgments}
This work was supported in part by the NYUAD Center for Quantum and Topological Systems (CQTS), funded by Tamkeen under the NYUAD Research Institute grant CG008, and the Center for Cyber Security (CCS), funded by Tamkeen under the NYUAD Research Institute Award G1104.

\bibliographystyle{IEEEtran}
\bibliography{refs}
\end{document}